\documentclass{article}
\usepackage{spconf,amsmath,graphicx,booktabs,cite,etoolbox,hyperref}
\usepackage[T1]{fontenc}
\usepackage[utf8]{inputenc}

\newcommand{\method}{SteerablePlex}
\newcommand{\bench}{SimIF-Bench}
\makeatletter
\apptocmd{\thebibliography}{\footnotesize\setlength{\itemsep}{0pt}\setlength{\parsep}{0pt}}{}{}
\makeatother

\title{STEERABLEPLEX: CAN WE STEER FULL-DUPLEX MODELS?}

\name{\parbox{0.96\textwidth}{\centering
Haolong Zheng\textsuperscript{1,*}, Maike Z\"ufle\textsuperscript{2},
Dominik Mach\'{a}\v{c}ek\textsuperscript{3}, Peter Pol\'{a}k\textsuperscript{3},
Xulin Fan\textsuperscript{1},\\
Xavier Sumba\textsuperscript{4}, Siyin Wang\textsuperscript{5},
Ond\v{r}ej Klejch\textsuperscript{6}, Mark Hasegawa-Johnson\textsuperscript{1,*}}}
\address{\parbox{0.96\textwidth}{\centering
\textsuperscript{1}University of Illinois Urbana-Champaign,
\textsuperscript{2}Karlsruhe Institute of Technology,\\
\textsuperscript{3}Charles University, \textsuperscript{4}Imperial College London,
\textsuperscript{5}Tsinghua University, \textsuperscript{6}University of Edinburgh\\
\textsuperscript{*}Corresponding authors: \{haolong2,jhasegaw\}@illinois.edu}}

\begin{document}
\maketitle

\begin{abstract}
Full-duplex speech models can listen and speak simultaneously, enabling natural interaction, but
become increasingly difficult to control as the conversation history grows. When used as user
simulators, this lack of control can cause them to deviate from prescribed scenarios and produce
unreliable evaluation outcomes. We introduce \bench{} (Simulator Instruction-Following Benchmark), which evaluates
whether a conversational model stays within a prescribed scenario and completes multiple goals in
the required order. The benchmark reveals that current open-source full-duplex models struggle to
follow such constraints. We then introduce a Group Reward-Decoupled Normalization Policy
Optimization (GDPO)-based training recipe that enables a full-duplex model to follow textual
instructions during an ongoing conversation while maintaining its turn-taking ability. By connecting the resulting \method{} to an asynchronous backend language model that
monitors the conversation and provides instructions when needed, we build a more controllable
full-duplex user simulator that follows multi-stage constraints more reliably than existing
open-source models and GPT-Realtime.
\end{abstract}


\section{Introduction}
\label{sec:intro}

Full-duplex spoken-dialogue models listen and speak simultaneously, generating speech or silence
while continuously encoding the partner's
audio~\cite{dgslm,moshi,syncllm,lslm,minicpmduplex,omniflatten,freezeomni,salmonnomni,salmduplex,personaplex,factor}. This capacity for
human-like conversation makes them especially promising as user simulators because they can reflect
how people actually interact, including overlap, interruptions, and backchannels. Naturalness alone,
however, is not sufficient. A valid simulator must adopt a prescribed persona, pursue a goal, and
create the situations that an evaluation is intended to probe~\cite{usersimsurvey,taubench,mehri2026goal}. It may need to
initiate a request, withhold information, or correct itself. Failure to follow
the scenario silently changes the test: the target agent
is evaluated on a different interaction, mixing simulator errors with the agent's ability.

For example, Full-Duplex-Bench v2~\cite{fdbv2} evaluates the task-specific capabilities of
full-duplex models by simulating a conversation between an \emph{examiner} and an
\emph{examinee}. In its original setup, GPT-Realtime~\cite{gptrealtime} serves as the examiner,
receiving a multi-stage script and steering the conversation to assess the examinee's task
performance. However, failure by the examiner to follow its script weakens the resulting assessment.
We therefore introduce \bench{} (Simulator Instruction-Following Benchmark) to quantify this
failure mode by measuring the examiner's topic adherence, naturalness, stage completion, and stage
order during interaction. We find that GPT-Realtime follows most individual instructions but remains
unreliable at completing the full sequence in order; current open-source full-duplex models exhibit
a larger gap. These results highlight the need for stronger control of full-duplex models.

Unlike a turn-based system, a full-duplex model continuously generates speech while listening and has no
clear boundary at which to adjust its behavior. Existing systems rely mainly on a system
prompt~\cite{personaplex,factor}, but this control can weaken as dialogue progresses: prior work
finds that system-prompt adherence drifts as conversation history
grows~\cite{instructionstability,laban2026lost}. Tight latency budgets also limit model size and its
capacity to execute long-horizon plans~\cite{moshirag}. Together, these limitations lead us to ask:

\begin{center}
\emph{Can we steer full-duplex models at runtime?}
\end{center}

We introduce \method{}, a full-duplex model that accepts instructions during a conversation, an
ability we call \textbf{online instruction following}. To enable this, we introduce a
special-token-delimited injection interface that inserts a control instruction into the running
model-text stream, and use GDPO~\cite{gdpo} to jointly optimize instruction following, response
naturalness, and turn taking. Inspired by KAME~\cite{kame} and MoshiRAG~\cite{moshirag}, we propose
an actor--director system that connects an asynchronous text-LLM director to \method{}. The director
consumes streaming ASR and model-text context, tracks the predefined task, and issues an instruction
only when needed. Together, these components yield a more controllable user simulator that follows
multi-stage task constraints more reliably than the evaluated baselines.

\begin{figure*}[t]
  \centering
  \includegraphics[width=\textwidth]{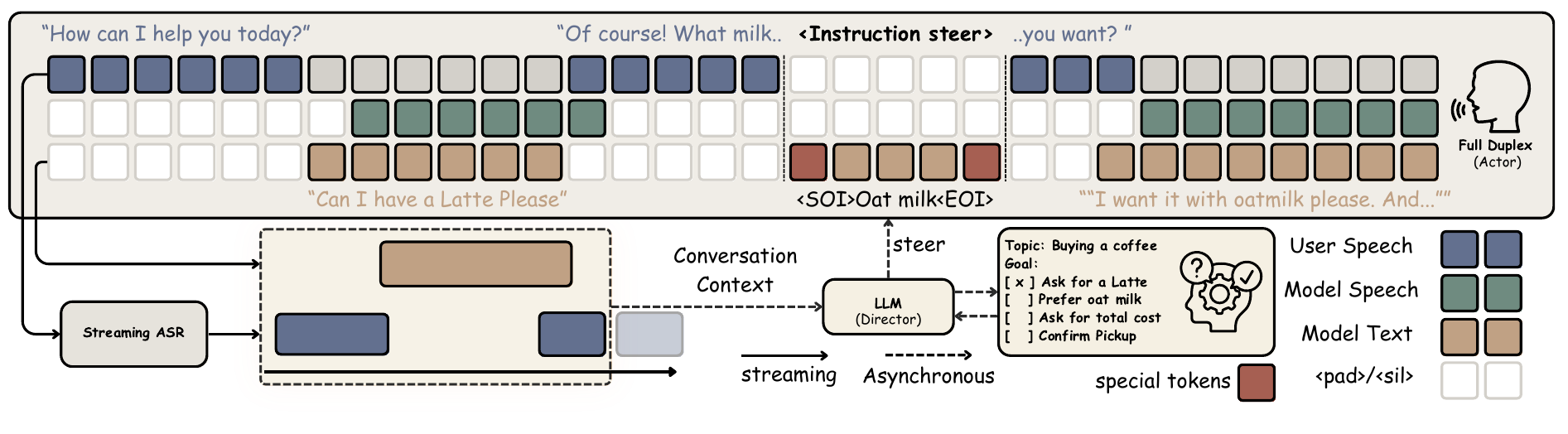}
  \caption{Illustration of the proposed Actor-Director system.
  \method{} continuously listens and speaks while the director monitors the conversation.
  When steering is needed, the director injects an instruction into the model-text stream.}
  \label{fig:system}
\end{figure*}

\begin{figure*}[t]
  \centering
  \includegraphics[width=\textwidth]{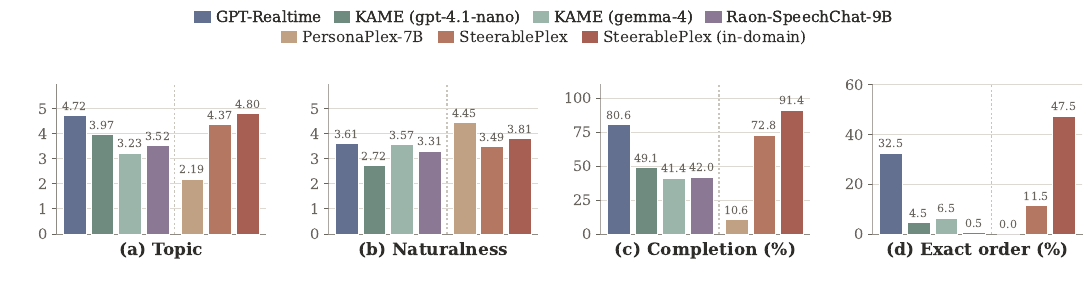}
  \caption{\bench{} results over 200 tasks. Panels report topic and naturalness (1--5),
  either-speaker goal completion, and examiner-only exact ordered completion. Both \method{}
  variants use the same actor--director harness.}
  \label{fig:fdif}
\end{figure*}


\section{\bench{}}
A reliable user simulator must do more than mention the requested content: it should complete
predefined stages in the prescribed order, remain focused on the target scenario despite unexpected
partner behavior, and steer the interaction without sacrificing conversational naturalness.
\bench{} evaluates these complementary requirements. It reuses the 200 staged scenarios and
interactive simulation setup of Full-Duplex-Bench v2~\cite{fdbv2}, but evaluates the examiner while
fixing PersonaPlex as the examinee. Calls end after 120\,s. Conversations are transcribed
and passed to an LLM judge~\cite{parakeet,llmjudge}. Three judging passes report topic adherence and naturalness
on 1--5 scales, conversation goal completion, and exact four-goal order. Categorical
decisions use majority vote, naturalness uses the median, and every decision must cite transcript
evidence.

\section{Methodology}
\label{sec:method}

To build a more controllable full-duplex user simulator, we propose the \textbf{Actor--Director}
architecture illustrated in Figure~\ref{fig:system}. The actor is a full-duplex model that runs
continuously as the front end, handling speech generation, listening, and turn taking. An asynchronous
text-LLM director tracks the task, conversation, and goals using context constructed from external
streaming ASR and the internal model-text stream. Every 2\,s, the director updates the goal state and
generates a candidate instruction together with an action\footnote{To avoid repeatedly steering the
front end with the same instruction, the director returns \texttt{send}, \texttt{wait}, or
\texttt{none}, indicating whether the instruction should be delivered, deferred, or omitted.}.
Below, we describe how we train the actor to follow director's instruction.

\subsection{Instruction Fusion}

\method{} builds on PersonaPlex~\cite{personaplex}, which consists of three synchronized streams:
user audio, model audio, and model text. The model-text stream internally predicts time-aligned
tokens for the model's own speech as a semantic prefix to the corresponding audio tokens~\cite{moshi}.
To encode a runtime instruction, we use the existing tokenizer and splice its tokens into the
model-text stream between special start- and end-of-instruction tokens. During these inserted steps,
both audio streams receive silence tokens and conversation time does not advance. This design
leverages the pretrained model's semantic understanding of the text stream, making it easier to
learn that a delimited text span is an instruction than to learn a new control modality.\footnote{We also observe
that the pretrained model can follow an injected instruction when it is phrased as a short
imperative, providing the initial capability needed for direct RL.}

\subsection{Data Preparation}
To prepare dialogue data with turn-level instructions, we begin with the Fisher dataset~\cite{cieri2004fisher} and use
Qwen3-32B~\cite{qwen3} to review each conversation turn and write a short, actionable instruction
when possible. For example, the instruction for ``Can I have a latte, please?'' is ``Ask for a
latte.'' From these instructions, we curate 9{,}417 RL examples, totaling 130.8 hours of rollout time.

To construct data closer to downstream simulator-style conversations, we follow the conversation-generation
pipeline of Behavior-SD~\cite{behaviorsd} to synthesize 2{,}000 overlapping-speech dialogues using
instructions derived from the 200 Full-Duplex-Bench v2 scenarios~\cite{fdbv2}. The resulting corpus
comprises 84.1 hours of stereo audio across 2{,}000 dialogues, of which 200 are held out.

\subsection{Instruction Following Training}
\noindent\textbf{GDPO-based Training:} GDPO extends Group Relative Policy Optimization (GRPO)~\cite{deepseekmath} to settings with multiple rewards~\cite{gdpo}.
GRPO samples a group of $G$ continuations for the same context and learns from their relative scalar
rewards. Directly combining heterogeneous rewards before this group normalization can allow a
high-variance component to dominate. GDPO instead normalizes each reward dimension $d$ independently
within the group as $z_{i,d}=(r_{i,d}-\mu_d)/\sigma_d$, sets undefined or flat dimensions to zero,
and then combines them as $\widetilde A_i=\sum_d w_dz_{i,d}$. Normalizing $\widetilde A_i$ once
more across the group gives the final advantage $A_i$. For a shared conversational context $h$, we
optimize
\begin{equation*}
\mathcal L_{\mathrm{RL}}(\theta)=\frac{1}{G}\sum_{i=1}^{G}\left[
-A_i\,\overline{\log\pi_\theta(y_i\mid h)}+
\beta\widehat D_{\mathrm{KL}}^{(i)}(\pi_\theta\Vert\pi_{\mathrm{ref}})\right],
\end{equation*}
where both terms average generated text and audio tokens and $\pi_{\mathrm{ref}}$ is the adapter-off
PersonaPlex policy. The KL term limits drift from the base conversational policy~\cite{instructgpt}.

\noindent\textbf{Rollout construction:} For Fisher, each rollout uses 30\,s of recorded two-party
context, inserts the instruction during the turn before its target response, and samples a 20\,s
model window while replaying the partner stream unchanged. Each synthetic dialogue contains four
ordered goals and yields one rollout per goal. Its prefix preserves the preceding conversation and
instructions, inserts the current instruction at its annotated turn, and uses the same sampling setup.

\begin{table*}[t]
\centering
\scriptsize
\caption{Results on Full-Duplex-Bench~\cite{fdbv1}.
$^*$ marks values reported in the original paper; unmarked rows are our runs.}
\label{tab:fdb}
\begin{tabular*}{\textwidth}{@{\extracolsep{\fill}}lcccccccccc@{}}
\toprule
& \multicolumn{2}{c}{Pause handling} & \multicolumn{3}{c}{Backchannel}
& \multicolumn{2}{c}{Smooth turn} & \multicolumn{3}{c}{Interruption} \\
\cmidrule(lr){2-3}\cmidrule(lr){4-6}\cmidrule(lr){7-8}\cmidrule(lr){9-11}
Model & Syn. $\downarrow$ & Cand. $\downarrow$ & TOR $\downarrow$ & Freq. $\uparrow$ & JSD $\downarrow$
& TOR $\uparrow$ & Lat. $\downarrow$ & TOR $\uparrow$ & Rating $\uparrow$ & Lat. $\downarrow$ \\
\midrule
dGSLM$^*$ & 0.934 & 0.935 & 0.691 & 0.015 & 0.934 & 0.975 & 0.352 & 0.917 & 0.201 & 2.531 \\
Moshi$^*$ & 0.985 & 0.980 & 1.000 & 0.001 & 0.957 & 0.941 & 0.265 & 1.000 & 0.765 & 0.257 \\
Freeze-Omni$^*$ & 0.642 & 0.481 & 0.636 & 0.001 & 0.997 & 0.336 & 0.953 & 0.867 & 3.615 & 1.409 \\
Gemini Live$^*$ & 0.255 & 0.310 & 0.091 & 0.012 & 0.896 & 0.655 & 1.301 & 0.891 & 3.376 & 1.183 \\
\midrule
KAME (gpt-4.1-nano) & 0.613 & 0.685 & 0.709 & 0.028 & 0.893 & 0.966 & 0.205 & 0.845 & 3.083 & 0.441 \\
KAME (gemma-4) & 0.620 & 0.718 & 0.727 & 0.025 & 0.897 & 0.975 & 0.151 & 0.820 & 3.521 & 0.356 \\
Raon-SpeechChat-9B & 0.650 & 0.880 & 0.473 & 0.171 & 0.694 & 1.000 & 0.017 & 1.000 & 3.791 & 0.236 \\
\midrule
PersonaPlex & 0.642 & 0.718 & 0.418 & 0.176 & 0.700 & 0.983 & 0.020 & 0.935 & 4.684 & 0.214 \\
\method{} & 0.394 & 0.500 & 0.218 & 0.113 & 0.750 & 1.000 & 0.156 & 1.000 & 4.790 & 0.264 \\
\method{} (in-domain) & 0.365 & 0.417 & 0.327 & 0.118 & 0.724 & 1.000 & 0.203 & 1.000 & 4.725 & 0.268 \\
\bottomrule
\end{tabular*}
\end{table*}

\noindent\textbf{Reward design:} Goal completion is the natural starting reward for instruction
following, but optimizing it alone admits two shortcuts: the model can blurt out the requested
content without integrating it into the conversation, or keep speaking over the partner to create
more opportunities for completion. We therefore complement goal completion with naturalness and
turn-taking rewards.

\emph{Goal completion} measures whether the current goal is actually fulfilled. Using only sampled
model speech produced after the corresponding instruction arrives, an LLM judge assigns rollout
$i$ a completion score $s_i$ of none, partial, or full.

\emph{Conversation naturalness}~\cite{arora2026rlaif} is scored once over the same sampled
continuation, encouraging the model to realize the requested behavior as a contextually appropriate
contribution rather than inserting it bluntly. The same judge assigns the window-level verdict
$q_i$: a score of one requires coherent speech from the model's own side of the conversation,
whereas self-answering, voicing the partner, or acting as an external assistant receives zero.

\begin{equation*}
r_i^{\mathrm{goal}}=s_i,\ s_i\in\{0,\tfrac12,1\},\qquad
r_i^{\mathrm{nat}}=q_i,\ q_i\in\{0,1\}.
\end{equation*}
\emph{Turn taking} preserves the base model's interactive behavior and prevents reward hacking in
which the model holds the floor or talks over the partner. Inspired by prior
work~\cite{interactivity,duplexpo}, we compute four VAD rewards from the sampled model, partner replay,
and reference continuation. We merge segments
separated by $<0.3$\,s and drop those shorter than 0.1\,s. Backchannels last at most 1\,s and overlap
partner speech by at least 50\%; $M_i$ denotes the remaining model speech and $U$ partner speech.
The reference labels a partner-silence gap as a yield, $y\in\mathcal Y$, if the reference takes the
floor, or a pause, $p\in\mathcal P$, if the partner resumes. We compute
\begin{equation*}
\begin{aligned}
r_i^{\mathrm{ovl}}&=-\left[\frac{|M_i\cap U|}{|M_i|}-0.10\right]_+, &
r_i^{\mathrm{take}}&=-\frac{1}{|\mathcal P|}\sum_{p\in\mathcal P}\chi_i(p),\\
r_i^{\mathrm{lat}}&=-\frac{1}{|\mathcal Y|}\sum_{y\in\mathcal Y}\Delta_i(y), &
r_i^{\mathrm{bc}}&=\operatorname{F1}(B_i,B^{\mathrm{ref}}),
\end{aligned}
\end{equation*}
Here $|\cdot|$ denotes duration and $[x]_+=\max(0,x)$. $\chi_i(p)=1$ when a model utterance longer than 1\,s begins in pause $p$, with a 0.3\,s
anticipation allowance; $\Delta_i(y)$ is the delay to the first model utterance of at least 1\,s,
capped at 5\,s, with 5\,s for no response. Backchannels $B_i$ and $B^{\mathrm{ref}}$ are greedily
onset-matched within $\pm1$\,s. All six rewards are higher-is-better. A turn-taking component is
omitted when its required event is absent; the reference continuation
identifies timing opportunities rather than prescribing what the model should say.

\section{Experiment}
\subsection{Experimental Setup}

Both variants initialize from PersonaPlex and train LoRA adapters~\cite{lora} (temporal/depth ranks 128/32;
scaling 2.0) against the adapter-off KL reference. GDPO uses $G=8$; audio/text sampling uses
temperatures 0.8/0.7 and top-$k$ 250/25. Both use learning rate $10^{-5}$: task-independent training
adapts $\beta$ from 0.05 toward KL target 0.3, while in-domain training fixes $\beta=0.01$.
Parakeet-TDT-0.6B-v2~\cite{parakeet,tdt} transcribes speech, and Silero VAD~\cite{silerovad} (threshold 0.5) extracts activity.
Qwen3.8-27B~\cite{qwen38} judges rewards and directs the actor.

\subsection{Baseline}

We compare GPT-Realtime~\cite{gptrealtime}, two KAME configurations~\cite{kame,gemma4}, and
Raon-SpeechChat-9B~\cite{raon} with PersonaPlex and our two RL variants. Figure~\ref{fig:fdif}
reports all 200 scenarios. GPT-Realtime is the strongest baseline. Compared with PersonaPlex, it trades some conversational
naturalness for stronger task fulfillment, yet still frequently fails to complete the goals in the
prescribed order. The evaluated open-source models lag further in completion and exact order.
KAME is the most closely related baseline: its full-duplex front end is trained to realize and
continue responses from a back-end LLM, demonstrating that a separate model can steer a
full-duplex front end. This improves completion over unsteered PersonaPlex but sacrifices
naturalness. Raon, a recent open-source model at the time of evaluation, likewise highlights the
difficulty of completing all goals in order under dynamic interaction.

PersonaPlex, our starting point, achieves the highest naturalness but the lowest completion and
exact-order scores. This pattern suggests that a model trained to react naturally to its partner can
still drift far from a prescribed scenario. Task-independent \method{} closes much of this gap and
outperforms the other open-source systems while maintaining a better completion--naturalness
balance. This result shows that the full-duplex front end can follow runtime instructions supplied
by a back-end model. However, its low exact-order success shows that it does not yet reliably
coordinate multiple instructions over a conversation. Because training uses single-instruction
examples, this result is consistent with limited transfer to multi-instruction streams. In contrast, structurally aligned training exposes \method{} (in-domain) to multiple ordered
instructions within each dialogue. It surpasses GPT-Realtime in both completion and exact order,
showing that matching the multi-instruction structure improves long-horizon control. This advantage
also holds on the 20 scenarios excluded from training.

\subsection{Turn-Taking Ability}

We follow the Full-Duplex-Bench~\cite{fdbv1} evaluation setup. Table~\ref{tab:fdb} includes its original
baselines and compares the released KAME and Raon checkpoints with the PersonaPlex.
KAME continuously steers its full-duplex front end with responses from a back-end model. This intervention substantially degrades all
three backchannel measures and reduces interruption success and quality, making the model harder to
interrupt. These results suggest that constant steering can weaken the front end's native
turn-taking policy.

In contrast, both \method{} variants are trained with explicit turn-taking rewards which helps preserve, and in several dimensions improving, the base
model's interaction ability. Relative to
PersonaPlex, they improve pause handling, backchannel TOR, smooth-turn TOR, interruption TOR, and
interruption rating, yielding gains on six of the ten reported metrics; their backchannel
distribution also remains close to the base model's. The remaining regressions are concentrated in
response latency and backchannel frequency.

\section{Conclusion}
We introduce \bench{} to evaluate instruction following in dynamic full-duplex conversations and
\method{}, a full-duplex front end trained with GDPO to follow runtime text instructions while
preserving turn taking. Coupled with an asynchronous director, \method{} provides a path toward controllable,
scenario-specific full-duplex user simulation.

\clearpage
\section{Acknowledgment}
This work was done in part at the 2026 Jelinek Memorial Summer Workshop on Speech and Language
Technology and was supported by NSF CCRI Grant No. 2120435, Google DeepMind, the JHU Amazon
Initiative for Interactive Artificial Intelligence, the JHU Human Language Technology Center of
Excellence, and the Association for Computational Linguistics.
This work used the Delta system at the National Center for Supercomputing Applications through
allocation bhxf-delta-gpu from the Advanced Cyberinfrastructure Coordination Ecosystem: Services
\& Support (ACCESS) program, which is supported by National Science Foundation grants \#2138259,
\#2138286, \#2138307, \#2137603, and \#2138296.

\bibliographystyle{IEEEbib}
\bibliography{refs}

\end{document}